# Remarkable Reduction of Thermal Conductivity in Silicon Nanotubes

Jie Chen,[1] Gang Zhang,[2,*] and Baowen Li[1,3]

[1]Department of Physics and Centre for Computational Science and Engineering, National University of Singapore, Singapore 117546, Singapore

[2] Key Laboratory for the Physics and Chemistry of Nanodevices and Department of Electronics, Peking University, Beijing 100871, People's Republic of China

[3]NUS Graduate School for Integrative Sciences and Engineering, Singapore 117456, Singapore

## Abstract

We propose to reduce the thermal conductivity of silicon nanowires (SiNWs) by introducing small hole at the centre, i.e. construct silicon nanotube (SiNT) structures. Our numerical results demonstrate that a very small hole (only 1% reduction in cross section area) can induce a 35% reduction in room temperature thermal conductivity. Moreover, with the same cross section area, thermal conductivity of SiNT is only about 33% of that of SiNW at room temperature. The spatial distribution of vibrational energy reveals that localization modes are concentrated on the inner and outer surfaces of SiNTs. The enhanced surface-to-volume ratio in SiNTs reduces the percentage of delocalized modes, which is believed to be responsible for the reduction of thermal conductivity. Our study suggests SiNT is a promising thermoelectric material with low thermal conductivity.

*E-mail: zhanggang@pku.edu.cn

Thermoelectric (TE) materials can provide electricity when subjected to a temperature gradient, or provide cooling performance when electrical current through it. They have the advantages of lightweight, environmentally benign and without moving parts. The efficiency of TE materials can be characterized by the dimensionless thermoelectric figure of merit $ZT = S^2\sigma T/\kappa$, where $S$, $\sigma$, $T$, and $\kappa$ are the Seebeck coefficient, electrical conductivity, absolute temperature, and thermal conductivity, respectively. As a result, materials with low thermal conductivity are highly desirable in order to achieve high ZT. Recently, experimental [1,2] and theoretical [3-5] efforts have demonstrated high thermoelectric performance of silicon nanowires (SiNWs), due to the low thermal conductivity observed in this low-dimensional material [6].

Thermal conductivity of nanoscale materials is quite different from that of bulk materials. For instance, due to the high surface-to-volume ratio (SVR) and the boundary scattering, thermal conductivity of SiNWs is about 2 orders of magnitude smaller than that of bulk crystal [7]. The low thermal conductivity of SiNWs is of particular interest for thermoelectric application. To further reduce the thermal conductivity, it has been suggested that random scattering of phonon is an efficient approach to reduce thermal conductivity [8, 9] and increase thermoelectric *ZT* correspondingly [10]. In addition to the random scattering of phonon, the surface scattering is another way to reduce thermal conductivity. It has been shown that surface roughness can decrease the phonon mean free path and in turn reduce the thermal conductivity [11].

In this Letter, we propose to reduce thermal conductivity further obviously by introducing more surface scattering: making SiNWs hollow to create inner surface, i.e. silicon nanotubes (SiNTs). Our numerical results demonstrate that only 1% reduction in cross section area (a very small hole) can induce 35% decrease of thermal conductivity at room temperature. Moreover, thermal conductivity of SiNTs decreases further with the increase of the size of the hole. The energy spatial distribution analysis demonstrates that localization mode on the surface is responsible for the low thermal conductivity in SiNTs.

We use equilibrium molecular dynamics (EMD) simulations to study the thermal conductivity of SiNTs and SiNWs along (100) with different cross section area. Here we set longitudinal direction along $x$ axis, and atoms in the same layers means they have the same $x$ coordinate. The atomic structure of NW is initially constructed from diamond structured bulk silicon, with $N_X$, $N_Y$, $N_Z$ unit cells in x, y and z direction, respectively. Then some central atoms in SiNWs are removed to create the SiNTs structures. The cross section of the hollow region is rectangular. The center of the rectangle is located near the geometric center of NW and the size is controlled by two parameters $L_y$ and $L_z$, which means there are ($2*L_y+1$) and ($2*L_z+1$) layers of silicon atoms are removed away in y and z direction, respectively. Fig. 1 shows a typical cross sectional view of SiNTs with $L_y=L_z=3$ and $N_Y=N_Z=5$. Here the SiNWs used to construct the NTs have a cross section area of 7.37 nm$^2$ ($N_Y=N_Z=5$). With adjustable $L_y$ and $L_z$ (from 1 to 6), the cross section area of SiNTs varies from 4.72 nm$^2$ to 7.30 nm$^2$.

In our simulations, Stillinger-Weber (SW) potential [12] is used to derive the force term as it can accurately describe elastic properties and thermal expansion coefficients [12-14]. In SW potential, it consists of a two-body term and a three-body term that can stabilize the diamond structure of silicon. Numerically, velocity Verlet algorithm is employed to integrate Newton's equations of motion, and each MD step is set as 0.8fs. Since quantum effect on thermal conductivity of SiNWs is quite small at room temperature [9], we do not adapt quantum correction in our study as we mainly concentrate on the geometric effect on thermal conductivity above room temperature.

In EMD simulations, thermal conductivity in longitudinal direction (along x axis) is calculated from Green-Kubo formula [15]

$$\kappa = \frac{1}{k_B T^2 V} \int_0^\infty dt \langle J_x(0) J_x(t) \rangle, \tag{1}$$

where $k_B$ is the Boltzmann constant, $V$ is the system volume, $T$ is the system temperature, $J_x(t)$ is the heat current along $x$ axis, and the angular bracket denotes an ensemble average. The heat current is defined as [16]

$$\vec{J}(t) = \sum_i \vec{v}_i \varepsilon_i + \frac{1}{2} \sum_{ij\ i \neq j} \vec{r}_{ij} (\vec{F}_{ij} \cdot \vec{v}_i) + \frac{1}{6} \sum_{\substack{ijk\ i \neq j \\ j \neq k}} (\vec{r}_{ij} + \vec{r}_{ik})(\vec{F}_{ijk} \cdot \vec{v}_i) \tag{2}$$

where $\vec{F}_{ij}$ and $\vec{F}_{ijk}$ denote the 2-body and 3-body force, respectively. A cubic super cell of $N_X \times N_Y \times N_Z$ unit cells is used in our simulation. Periodic boundary condition is applied in $x$ (longitudinal) direction, and free boundary condition is applied in other two directions, the atoms on the inner- and outer- surfaces of SiNTs and SiNWs. For each realization, all the atoms are initially placed at their equilibrium positions but have a random velocity according to Gaussian distribution. Canonical ensemble MD with Langevin heat reservoir first runs for $10^5$ steps to equilibrate the whole system at a given temperature. Then micro-canonical ensemble MD runs for another $3 \times 10^6$ steps (2.4 ns) and heat current is recorded at each step. After that, thermal conductivity is calculated according to Eq. (1). The final result is averaged over 6 realizations with different initial conditions.

Fig. 2(a) shows the time dependence of normalized heat current autocorrelation function (HCACF) for a typical realization in a 16×5×5 super cell at 300K (other realizations are similar). It shows a very rapid decay of HCACF at the beginning, followed by a long tail which has a much slower decay. This two-stage decaying characteristic of HCACF has been found in the study of various materials [16-18]. The rapid decay corresponds to the contribution from short wavelength phonons to thermal conductivity, while the slower decay corresponds to the contribution from long wavelength phonons [17, 18]. Furthermore, it is shown in Fig. 2(a) that HCACF decays to approximately zero within 100 ps (cut-off time), much shorter than the total simulation time of 2.4 ns. It has been checked that this is also true for the largest super cell size $N_X=20$ considered in our study. Therefore, the total simulation time of 2.4 ns is adequate for the present study.

Using the standard EMD approach in the integration of HCACF up to the cut-off time, we calculate the thermal conductivity of SiNWs according to Eq. (1). Fig. 2(b) shows the calculated thermal conductivity of SiNWs with a fixed cross section of 5×5 unit cells versus super cell length $N_X$ at 300K. Due to the periodic boundary condition, finite size effect exists in the calculated thermal conductivity when simulation domain is small [16, 17]. In our simulations, the thermal conductivity of SiNWs saturates to a constant when the super cell size $N_X \geq 16$ unit cells, in agreement with previous study [7]. Therefore, in the following part, we set $N_X=16$ in the longitudinal direction and study the thermal conductivity of SiNWs and SiNTs with different cross section area.

Fig. 3(a) shows the thermal conductivity of SiNWs and SiNTs versus cross section area at 300K. Here the cross section area of SiNT is defined as the area of corresponding SiNW minus the removed part. Even with a very small hole, $Ly=Lz=1$, the thermal conductivity decreases obviously, from $\kappa_{NW}=12.2\pm1.4$ W/mK to $\kappa_{NT}=8.0\pm1.1$ W/mK. In this case, only a 1% reduction in cross section area (from 7.37

nm$^2$ to 7.30 nm$^2$) induces the reduction of thermal conductivity of 35%. Fig. 3(b) shows the thermal conductivity of SiNWs (cross section area 7.37 nm$^2$) and SiNTs (cross section area 7.30 nm$^2$) at different temperature. Thermal conductivity of both SiNWs and SiNTs decreases with increasing of temperature, which is in general a consequence of the stronger anharmonic phonon-phonon scattering at higher temperature. Our results are in agreement with a recent study on the temperature dependent thermal conductivity of thin SiNWs [19]. In addition, when temperature increases, thermal conductivity of SiNWs decreases slightly faster than that of SiNTs does. This is because the creation of the hollow centre results in more localized phonon modes in SiNTs. It has been demonstrated that the introduction of localization modes can weaken the temperature dependence of thermal conductivity [20], which is also recently observed in the crystalline-core/amorphous-shell SiNWs [19]. So the reduction percentage in thermal conductivity slightly decreases from 35% at 300K to 29% at 1000K. However, even at high temperature as 1000K, the impact of small hole on thermal conductivity is still obvious, a 1% reduction in cross section area can induce a 29% reduction in thermal conductivity.

In addition, with increasing of $Ly$ and $Lz$, the cross section area decreases further, and a linear dependence of thermal conductivity on cross section area is observed, as shown in Figure 3(a). We also show the thermal conductivity of SiNW with $Ny=Nz=4$. It has the same cross section area (4.72 nm$^2$) as the SiNT with $Ny=Nz=5$ and $Ly=Lz=6$. It is clear that for SiNW, thermal conductivity increases with cross section area increases. This is because with the increase of size, more and more phonons are excited, which results in the increase of thermal conductivity. So the decrease of cross section area is one origin for the low thermal conductivity of SiNT. However, it is not the sole one. We can see that with the same cross section area, thermal conductivity of SiNTs is only about 33% of that of SiNWs ($\kappa_{NW}$=8.8±1.1 W/mK and $\kappa_{NT}$=2.9±0.5 W/mK). From below we will demonstrate this additional reduction is due to the localization of phonon states on the surface.

Due to the inner surface in SiNTs which partially destroys the original periodicity of SiNWs, phonon localization takes place on the surface in general. To understand the underlying physical mechanism of thermal conductivity reduction in SiNTs, we carry out a vibrational eigen-mode analysis on SiNWs and SiNTs. Mode localization can be quantitatively characterized by the participation ratio $P_\lambda$ defined for each eigen-mode $\lambda$ as [21]

$$P_\lambda^{-1} = N \sum_i \left( \sum_\alpha \varepsilon_{i\alpha,\lambda}^* \varepsilon_{i\alpha,\lambda} \right)^2, \qquad (3)$$

where $N$ is the total number of atoms, and $\varepsilon_{i\alpha,\lambda}$ is the α-th eigenvector component of eigen-mode λ for the i-th atom. The participation ratio measures the fraction of atoms participating in a given mode, and effectively indicates the localized modes with O(1/N) and delocalized modes with O(1). It can provide more detailed information about localization effect for each mode.

Fig. 4 compares the participation ratio (p-ratio) of each eigen-mode for SiNWs and SiNTs with the same cross section area (4.72 nm$^2$). It shows a reduction of p-ratio in SiNTs for both low frequency phonons and high frequency phonons, compared with SiNWs. Most of the eigen-modes in SiNWs have p-ratio greater than 0.5, showing characteristic of delocalized mode, while majority of the eigen-modes in SiNTs have p-ratio less than 0.5, showing characteristic of localized mode.

Although participation ratio can effectively describe mode localization in a quantitative manner, it does not provide information about the spatial distribution of a specific mode. To get a better physical picture about the localization modes, we also provide the local vibrational density of states (LVDOS) which is defined as [21,22]

$$D_i(\omega) = \sum_\lambda \sum_\alpha \varepsilon^*_{i\alpha,\lambda} \varepsilon_{i\alpha,\lambda} \delta(\omega - \omega_\lambda), \qquad (4)$$

where $i$ denotes the $i$-th atom that corresponds to a specific spatial location. Based on LVDOS, we define the spatial distribution of energy as:

$$\begin{aligned} E_i &= \sum_\omega (n+\frac{1}{2})\hbar\omega D_i(\omega) \\ &= \sum_\omega \sum_\lambda \sum_\alpha (n+\frac{1}{2})\hbar\omega \varepsilon^*_{i\alpha,\lambda} \varepsilon_{i\alpha,\lambda} \delta(\omega - \omega_\lambda) \end{aligned}, \qquad (5)$$

where $n$ is the phonon occupation number given by the Bose-Einstein distribution. Instead of looking at energy distribution for each mode, here we can obtain the total energy spatial distribution within a specified frequency regime.

In order to observe the spatial localization which is caused by the localized modes, the summation of $\omega$ in Eq. (5) only includes those modes with a relatively small p-ratio (e.g. 0.2). Fig. 5 shows the normalized energy distribution on the cross section (YZ) plane for SiNWs and SiNTs at 300K. The positions of the circles denote different locations on YZ plane. For those localized modes with p-ratio less than 0.2, it is clearly shown in Fig. 5(a) that the intensity of localized modes is almost zero in the centre of NW, while with finite value at the boundary. This demonstrates that the localization modes in SiNWs are distributed on the boundary (especially at the corner) of cross section plane, which corresponds to the outer surface of SiNWs. In addition, due to inner-surface introduced in SiNTs, energy localization also shows up around the hollow region as shown in Fig. 5(b). These results provide direct numerical evidence that localization takes place on the surface region. Alternatively, we also study the energy distribution of delocalized modes with a relatively large p-ratio (e.g. 0.6). For those delocalized modes with p-ratio greater than 0.6, majority of energy is distributed inside SiNWs and SiNTs as shown in Fig. 5(c)-(d), except for the hollow region in SiNTs. Therefore, from the spatial distribution of delocalized modes, we can conclude that energy is localized on the boundary (surface) region.

Since SiNTs in our simulation are constructed from SiNWs but have hollow interior, the density is approximately the same for these two structures. For SiNWs and SiNTs with the same cross section area and same length, the volume is the same. Therefore, the total number of atoms, thus the total number of eigen-modes, is the same for SiNTs and SiNWs under this condition. Compared with SiNWs, SiNTs have a larger surface area, which corresponds to a higher SVR. As a result, there are more modes localized on the surface, which increases the percentage of the localized modes to the total number of modes. This explains the overall reduction of p-ratio in SiNTs compared with SiNWs as shown in Fig. 4. In heat transport, the contribution to thermal conductivity mainly comes from the delocalized modes rather than the localized modes. Due to the enhanced SVR in SiNTs which induces more localized modes, the percentage of delocalized modes decreases, leading to a reduction of thermal conductivity in SiNTs compared with SiNWs.

To summarize, in this letter, we have proposed to reduce the thermal conductivity of SiNWs by introducing hollow interior at the centre, i.e. construct SiNT structures. The molecular dynamics simulations results demonstrate that at room temperature with only a small hole of 0.07 nm$^2$, (cross section area from 7.37 nm$^2$ to 7.30 nm$^2$), it can reduce thermal conductivity from $\kappa_{NW}$=12.2 W/mK to $\kappa_{NT}$=8.0 W/mK, which means a 1% reduction in cross section area induces a 35% reduction in thermal conductivity. Moreover, with the same cross section area, thermal conductivity of SiNTs is only about 33% of that of SiNWs at room temperature. Compared to SiNWs with the same cross section area, participation ratio in SiNTs decreases for both low frequency phonons and high frequency phonons. The spatial distribution of energy reveals that localization takes place on the surface region in both SiNWs and SiNTs. The enhanced surface-to-volume ratio in SiNTs reduces the percentage of delocalized modes, and thus lowers the thermal conductivity. Very recently, the similar SiNT structures have been fabricated experimentally by reductive decomposition of a silicon precursor in an alumina template and etching [23]. Our results suggest that SiNTs is a promising thermoelectric material by using reliable fabrication technology.

**Acknowledgement**

This work was supported in part by Grant R-144-000-222-646 from the National University of Singapore.

**Reference**

[1] A. I. Hochbaum, R. Chen, R. D. Delgado, W. Liang, E. C. Garnett, M. Najarian, A. Majumdar, and P. Yang, Nature **451**, 163 (2008).

[2] A. I. Boukai, Y. Bunimovich, J. T. Kheli, J-K Yu, W. A. Goddard III, and J. R. Heath, Nature **451**, 168 (2008).

[3] T. Vo, A. J. Williamson, V. Lordi, and G. Galli, Nano Lett. **8**, 1111 (2008).

[4] G. Zhang, Q. Zhang, C.-T. Bui, G.-Q. Lo, and B. Li, Appl. Phys. Lett. **94**, 213108 (2009).

[5] G. Zhang, Q.-X. Zhang, D. Kavitha, and G.-Q. Lo, Appl. Phys. Lett. **95**, 243104 (2009).

[6] D. Li, Y. Wu, P. Kim, L. Shi, P. Yang, and A. Majumdar, Appl. Phys. Lett. **83**, 2934 (2003).

[7] S. G. Volz and G. Chen, Appl. Phys. Lett. **75**, 2056 (1999).

[8] N. Yang, G. Zhang, and B. Li, Nano Lett. **8**, 276 (2008).

[9] J. Chen, G. Zhang, and B. Li, Appl. Phys. Lett. **95**, 073117 (2009).

[10] L. Shi, D. Yao, G. Zhang, and B. Li, Appl. Phys. Lett. **95**, 063102 (2009).

[11] D. Donadio, and G. Galli, Phys. Rev. Lett., **102**, 195901 (2009).

[12] F. H. Stillinger, and T. A. Weber, Phys. Rev. B **31**, 5262 (1985).

[13] M. D. Kluge, J. R. Ray, and A. Rahman, J. Chem. Phys. **85**, 4028 (1986).

[14] J. Q. Broughton, and X. P. Li, Phys. Rev. B **35**, 9120 (1987).

[15] R. Kubo, M. Toda, and N. Hashitsume, Statistical Physics (Berlin, Springer, 1985), Vol. 2.

[16] P. K. Schelling, S. R. Phillpot, and P. Keblinski, Phys. Rev. B **65**, 144306 (2002).

[17] J. Che, T. Cagin, W. Deng, and W. A. Goddard, J. Chem. Phys. **113**, 6888 (2000).

[18] A. J. H. McGaughey, and M. Kaviany, Int. J. Heat Mass Transfer **47**, 1783 (2004).

[19] D. Donadio, and G. Galli, Nano Lett. **10**, 847 (2010).

[20] J. L. Feldman, M. D. Kluge, P. B. Allen, and F. Wooten, Phys. Rev. B **48**, 12589 (1993).

[21] A. Bodapati, P. K. Schelling, S. R. Phillpot, and K. Keblinski, Phys. Rev. B **74**, 245207 (2006).

[22] J. L. Feldman, and N. Bernstein, Phys. Rev. B **70**, 235214 (2004).

[23] M. Park, M. Kim, J. Joo, K. Kim, J. Kim, S. Ahn, Y. Cui, and J. Cho, Nano Lett. **9**, 3844 (2009).

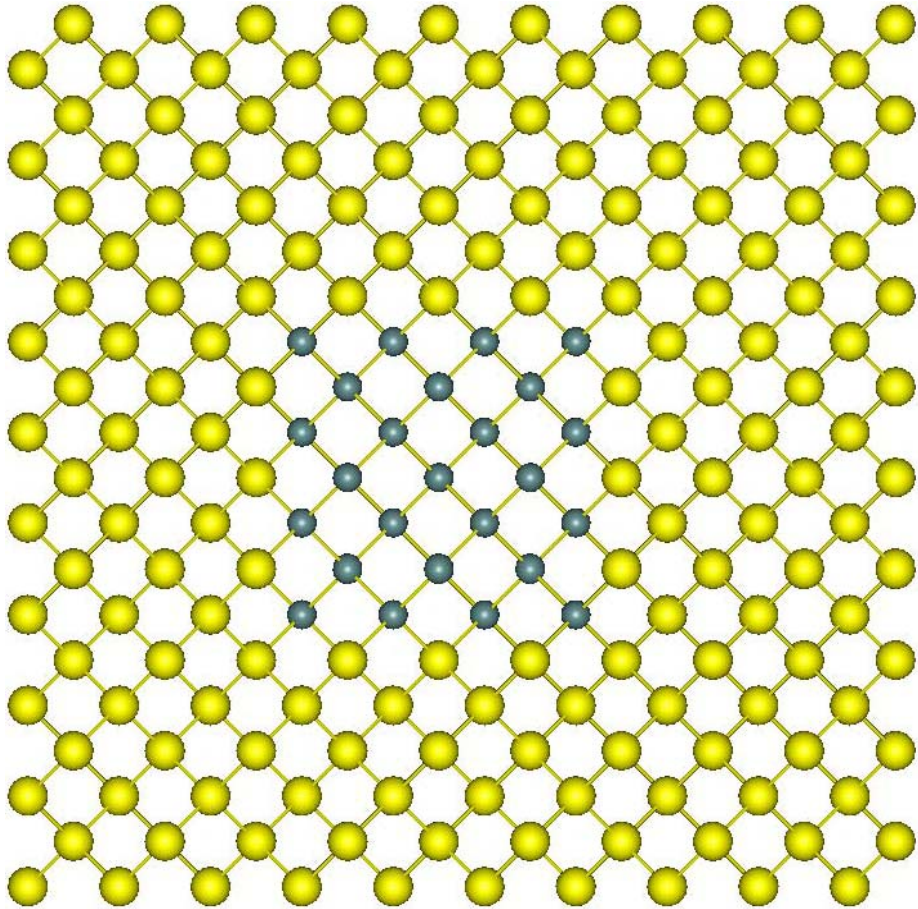

Figure 1. (color online) A typical cross sectional view of the SiNTs with $L_y=L_z=3$ and $N_Y=N_Z=5$ (see text). The central atoms in green denote the removed atoms.

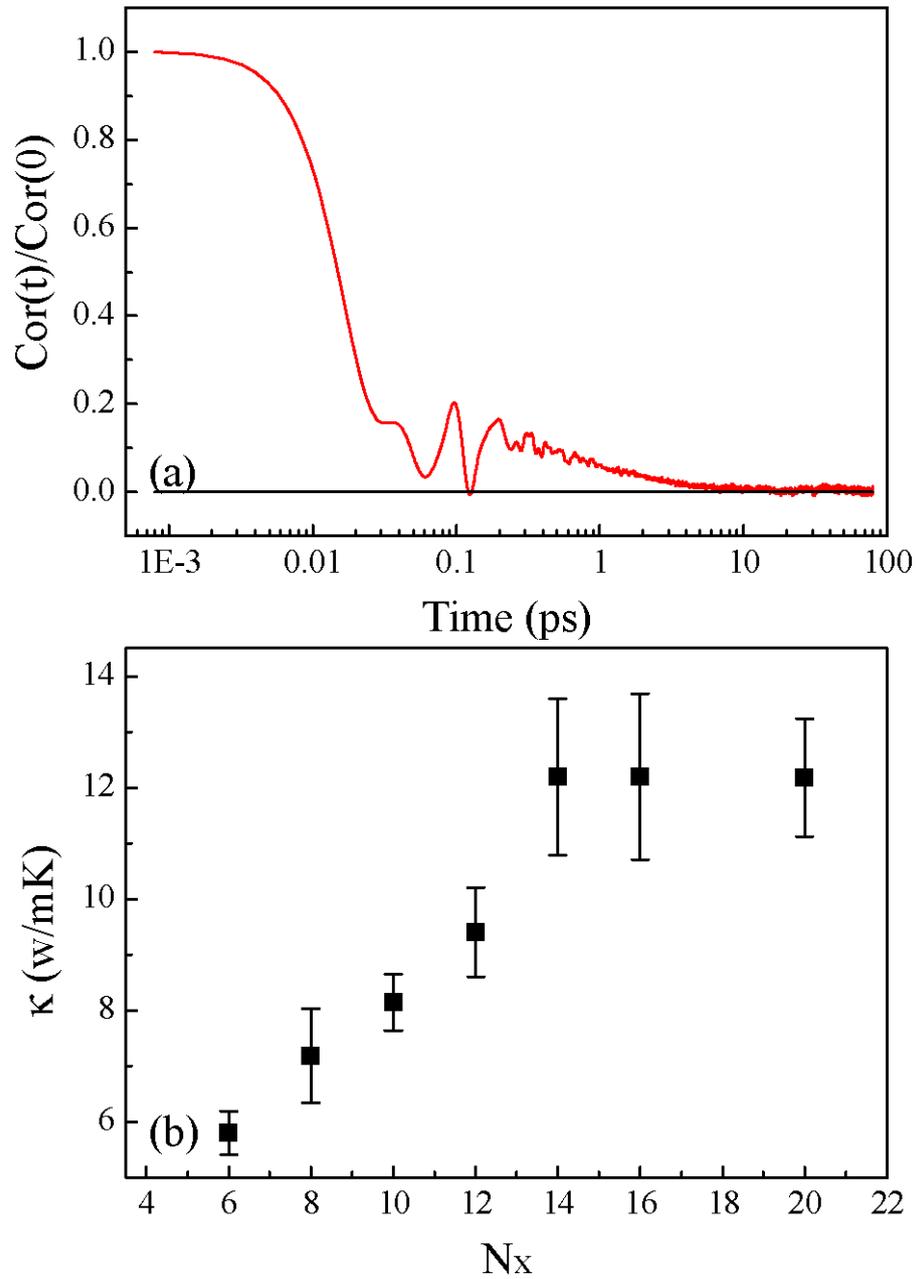

Figure 2. (color online) (a) Time dependence of normalized heat current autocorrelation function Cor(t)/Cor(0) (red line) for a typical realization in a 16×5×5 super cell at 300K. The black line draws the zero-axis for reference. (b) Thermal conductivity of SiNWs at 300K versus super cell size $N_X$ ($N_X$×5×5 unit cells).

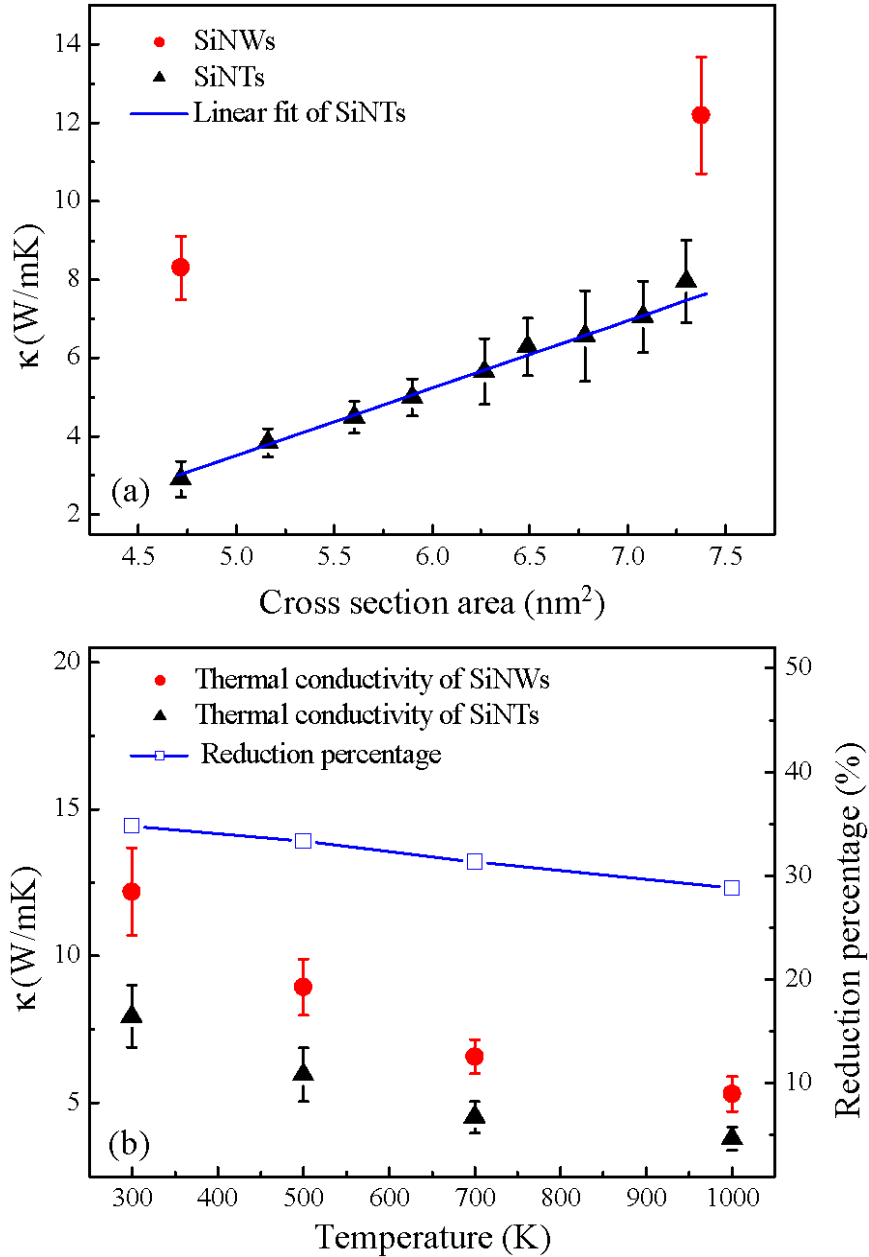

Figure 3. (color online) (a) Thermal conductivity of SiNWs and SiNTs versus cross section area at 300K. The red dot and black triangle denote SiNWs and SiNTs, respectively. The cross section areas for these two SiNWs are 7.37 nm$^2$ and 4.72 nm$^2$, respectively. The blue line is drawn to guide the eyes. (b) Thermal conductivity of SiNWs and SiNTs versus temperature. The red dot and black triangle denote SiNWs and SiNTs, respectively. The cross section areas for SiNWs and SiNTs are 7.37 nm$^2$ and 7.30 nm$^2$, respectively. The blue square draws the reduction percentage in thermal conductivity from SiNWs to SiNTs.

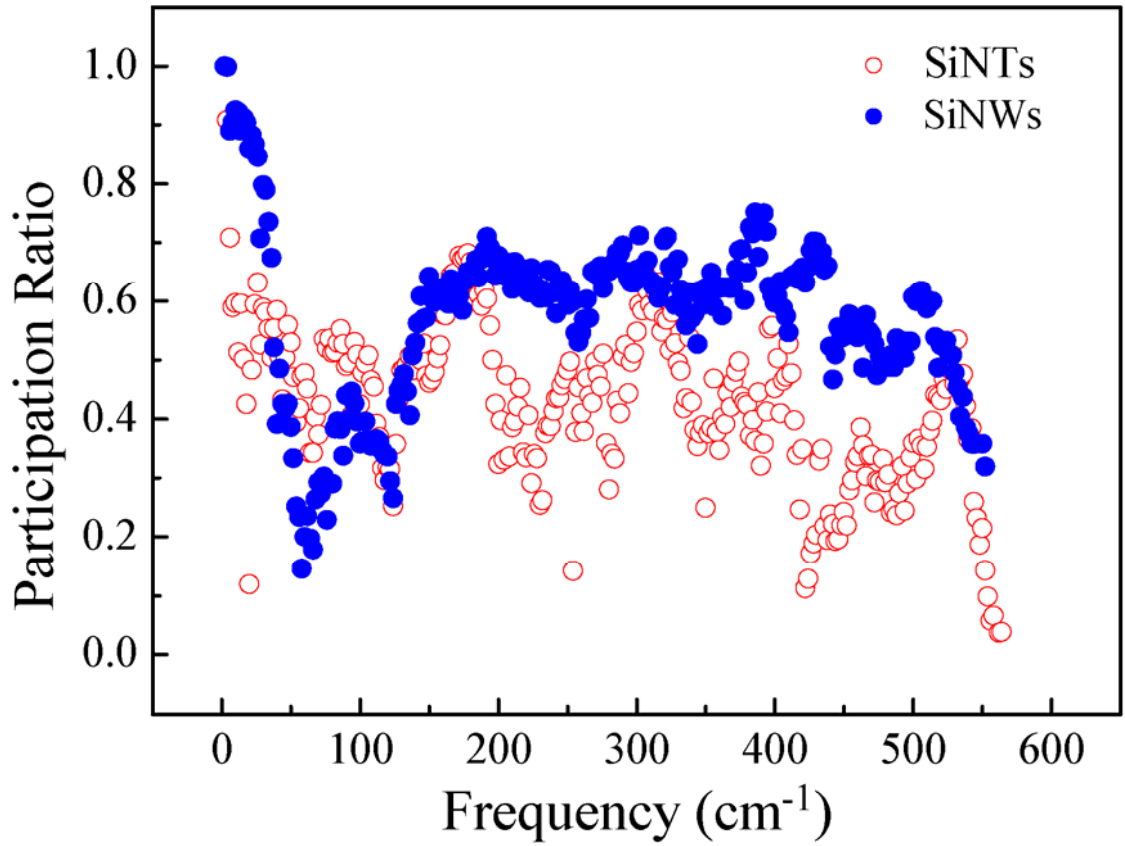

Figure 4. (color online) Participation ratio of each eigen-mode for SiNTs and SiNWs with the same cross section area of 4.72 nm$^2$. The hollow red circle and solid blue circle denote SiNTs and SiNWs, respectively.

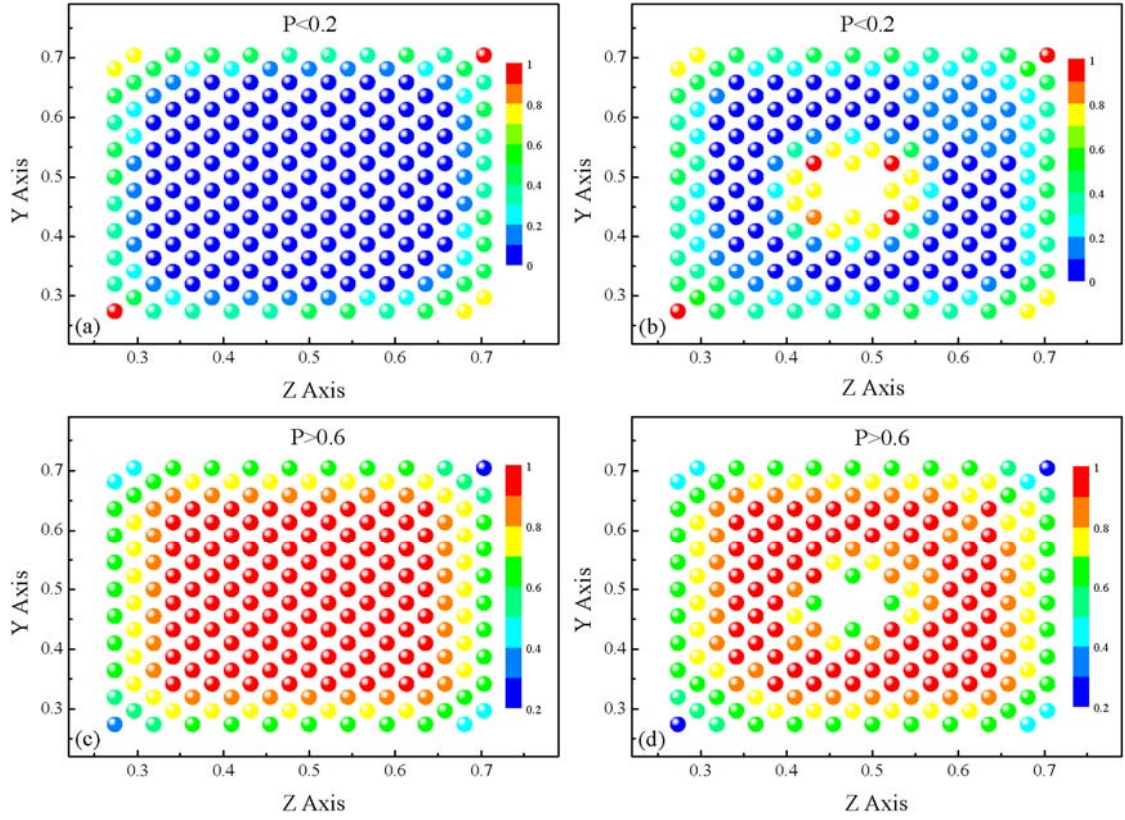

Figure 5. (color online) Normalized energy distribution on the cross section (YZ) plane for SiNWs and SiNTs at 300K. Positions of the circles denote the different locations on YZ plane, and intensity of the energy is depicted according to the color bar. P in the figure denotes participation ratio. (a) Energy distribution of SiNWs for modes with P<0.2; (b) Energy distribution of SiNTs for modes with P<0.2; (c) Energy distribution of SiNWs for modes with P>0.6; (d) Energy distribution of SiNTs for modes with P>0.6.